\documentclass[%
 aip,
 amsmath,amssymb,
 reprint,%
]{revtex4-1}

\usepackage{graphicx}
\usepackage{dcolumn}
\usepackage{bm}

\usepackage[utf8]{inputenc}
\usepackage[T1]{fontenc}
\usepackage{mathptmx}
\usepackage{etoolbox}
\usepackage{multirow}
\usepackage{colortbl}
\usepackage{xcolor} 
\newcommand\myred[1]{\textcolor{black}{#1}}  
\newcommand\mygreen[1]{\textcolor{black}{#1}} 
\newcommand{\voldash}{\makebox[1pt][l]{$-$}V}

\makeatletter
\def\@email#1#2{%
 \endgroup
 \patchcmd{\titleblock@produce}
  {\frontmatter@RRAPformat}
  {\frontmatter@RRAPformat{\produce@RRAP{*#1\href{mailto:#2}{#2}}}\frontmatter@RRAPformat}
  {}{}
}%
\makeatother
\begin{document}

\preprint{AIP/123-QED}

\title[]{Open inverted bell and bell formation during the washing of vials}

\author{Javed Mohd.}
\author{Amar Yadav}%
\author{Debopam Das}
\email{das@iitk.ac.in}
\affiliation{ 
Department of Aerospace Engineering, Indian Institute of Technology Kanpur, Kanpur 208016, India}


\date{\today}

\begin{abstract}
A range of fascinating flow features was observed while cleaning or filling the medicine or liquid vials in the kitchen sink by serendipity. Here, we present the formation of an open inverted bell and bell-shaped flow structures formed from the water sheet in a new geometric arrangement, hitherto unknown. When a laminar jet impinges on the surface of the liquid in the vial of marginally larger or similar diameter, an inverted open water bell is formed, which gradually changes into a flat water sheet to classical water bell as the flow rate is increased. The inverted water bell structures disintegrate by forming water ridges which finally break down into different sizes of droplets  
\end{abstract}

\maketitle

\section{\label{sec:intro}Introduction}
\begin{figure}
\includegraphics[width=0.45\textwidth]{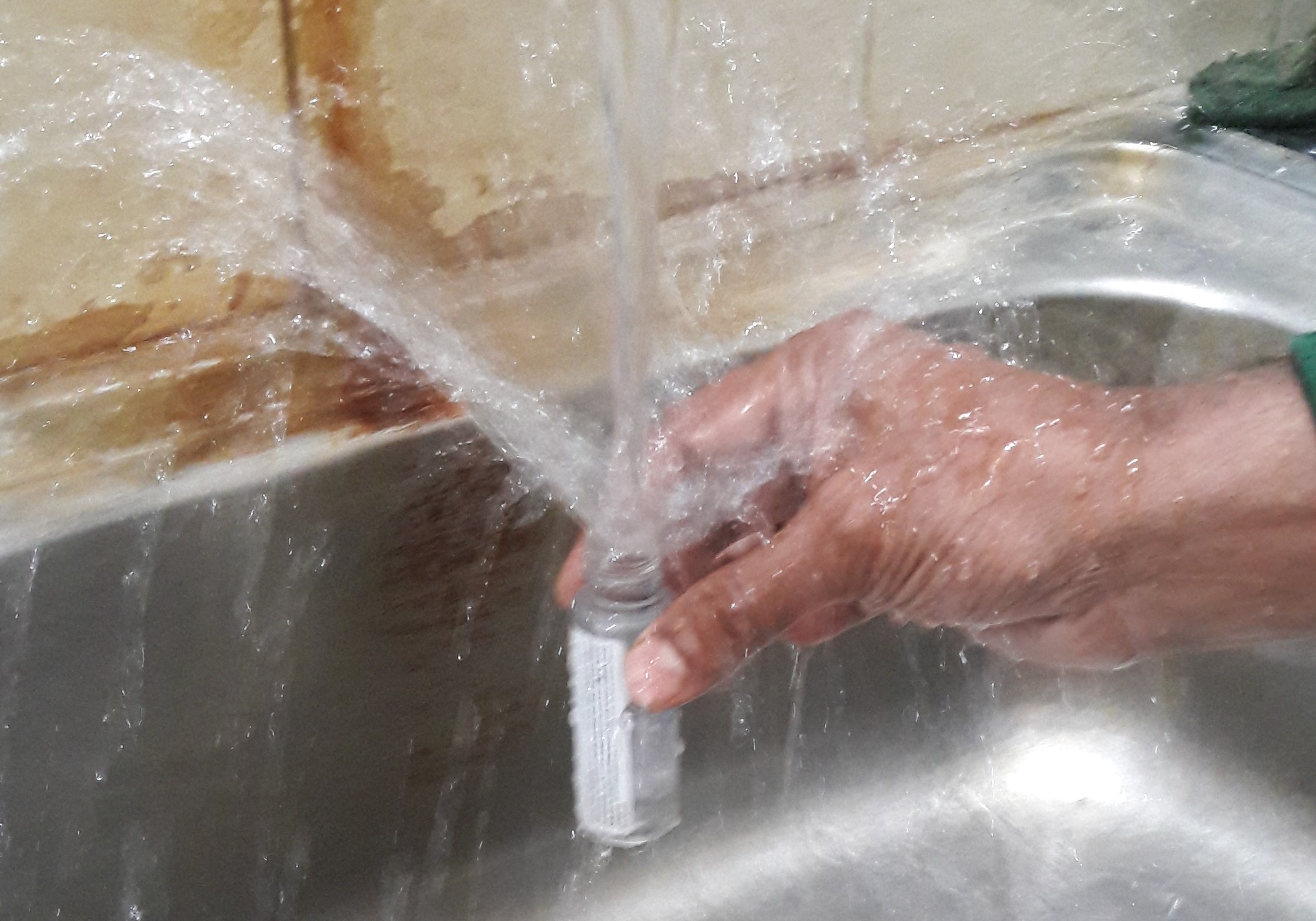}
\caption{\label{fig:introKitchen} A typical vial washing activity in kitchen sink}
\end{figure}

Many of the kitchen products as well as medicines are supplied in small to medium-sized glass vials.
A myriad of fascinating flow structures develop while washing/filling these vials (as in figure \ref{fig:introKitchen}) in the kitchen sink. In a similar vial washing activity, it was noted that the water while gushing out of the \mygreen{vial} forms spectacular\mygreen{, steady} inverted umbrella-like water sheets. The shape can be described as an open inverted water bell (see figures \ref{fig:case1} {\color{black}(a) (Multimedia view) and \ref{fig:case2} (b) (Multimedia view))}. Even, the formation of classical closed water bell \cite{savart1833memoire_a, boussinesq1869theorie, hopwood1952water, clanet2001dynamics, Taylor1959_a} has been observed {\color{black}(figures \ref{fig:case1} (c) (Multimedia view) and \ref{fig:case4} (c) (Multimedia view))} in this new arrangement which is generally formed when the liquid jet impinges on a solid
surface. The shape and size of this inverted umbrella-like
water sheets change by increasing the flow rate or changing
the vial size.   

To understand and characterize the inverted bell, laboratory-scale experiments were designed and performed in a controlled environment. Nozzles of different diameters, mimicking the kitchen faucets, and cylindrical tubes mimicking the vials were used in the experiments. The shape and stability of water sheets are prominently governed by the interplay of surface tension forces, inertial forces and gravity \cite{boussinesq1869theorie, lance1953water, Taylor1959_a, baird1962annular, wegener1964surface}. Several studies were conducted in the past to model the behavior of water sheet structures; clustered around the study concerning a particular water sheet shape known as the water bell as summarized by Clanet \cite{clanet2007waterbells}. A water bell is formed when a freely moving water film encloses and traps an amount of air inside it.  Almost all of these earlier studies produced the water bell using a solid disk deflector that provides the no-penetration boundary condition at the deflector wall. A no-penetration boundary condition is also provided by the liquid in a few cases reported such as by \mygreen{Savart} \cite{savart1833memoire_d} in case of head on collision of two jets of different diameter which generated a water bell. Paramati and Tirumkudulu \cite{paramati2016open} used two head-on colliding jets in a uniform velocity air to produce open water bells. Marmottant et al. \cite{marmottant2000transient} used a ring around the deflector plate to force the sheet upward. However, the no-penetration boundary condition is still being supplied by the solid deflector disk. Transient, inverted open water bells have been reported in the form of an ejecta sheet when a droplet is impinging in a pool of the same liquid \cite{thoroddsen2002ejecta, engel1966crater}. However, \mygreen{there exist no published studies examining a continuous jet impinging on the surface of the liquid in the vial and forming such steady structures}. Therefore, to the best of the authors' knowledge, this is the first time that the water bells formed by the impingement of a laminar jet on the surface of the liquid in the cylindrical vial are being reported. Interestingly, this arrangement forms upward inverted bell to horizontal water sheet to downward classical bell depending on the geometric and flow parameters.

The earliest account of water bells could be traced to the experimental study of Savart \cite{savart1833memoire_a,savart1833memoire_b, savart1833memoire_c, savart1833memoire_d}. Boussinesq \cite{boussinesq1869theorie} provided mathematical modeling for the experimental shape obtained by Savart. Bond \cite{bond1935surface} and Buchwald and K''oning, \cite{buchwald1936dynamic} used the radius of a horizontal sheet formed due to head-on collision of two liquid jets to calculate the surface tension of water. 
Taylor \cite{Taylor1959_a} used a conical or planer deflector with a horizontal nozzle to generate the water bell. From the shape of the water bells, he found out that gravity does not affect the sheet shape significantly. In the follow-up investigations \cite{Taylor1959_b,Taylor1959_c} he investigated the waves forming on the surface of water bells and break-up dynamics of these sheets. {\color{black} Kolinski et al. \cite{PhysRevFluids.2.042401} } investigated the growth of harmonic perturbations on the water bell. They perturbed the bell by changing the height of the collar attached to the flat impactor which resulted in the ejection angle variation around its mean value.

Marmottant et al.\cite{marmottant2000transient} used a solid deflector with a \mygreen{adjustable} ring for experimental setup to measure the transient characteristics of the horizontal water bell. Clanet \cite{clanet2001dynamics, clanet2000stability} used a solid disk impactor and vertically discharging nozzle similar to Savart \cite{savart1833memoire_a} to study the dynamics and stability of the water bells. He conducted a parametric experimental investigation based on the diameter ratio, $X=D_i / D_o$ where $D_i$ was the diameter of solid impactor disk and $D_o$ was the diameter of the impinging jet. The regime for which he observed the formation of water bells was defined in terms of X as a function of Webber number (We) and Reynolds number (Re). In the follow-up study Clanet and Villermaux \cite{clanet2002life}, and {\color{black}Villermaux and Almarcha\cite{PhysRevFluids.1.041902}} studied the atomization and fragmentation of water sheet formed. They used a ring-type lip to produce water sheets with an ejection angle greater than $\pi/2$. {\color{black}Lhuissier and Villermaux \cite{lhuissier2012crumpled}} used a similar set-up to investigate the formation of crumpled water bell. {\color{black}Aristoff et al. \cite{AristoffBush2006}} used an adjustable lip adjoining the impactor plate to produce different bell shapes using the glycerol-water solution. {\color{black}Buckingham and Bush \cite{buckingham2001fluid}} has also reported the formation of polygonal bells. {\color{black}Speirs et al. \cite{PhysRevFluids.3.100504}} suddenly released a mass of water held in a polycarbonate tube and reported the formation of an upside-down water bell that also resembled the shape of the fluted champagne glass.

Instead of a liquid jet impinging on a circular deflector disk, Jameson et al. \cite{jameson2008water,jameson2010water} used a liquid jet directed vertically upwards to a large deflector plate and reported the formation of a new type of water bell. Theoretical modeling was provided in a companion paper by Button et al. \cite{button2010water}.    

In the current study, we performed experiments using a vial instead of an impactor disk or a ring-shaped lip. The observed water sheet structures are presented by means of four cases having different flow configurations based on the nozzle diameter, vial diameter, and distance between nozzle and vial. In all four cases that we present, the nozzle and vial diameter is fixed, in each case, whereas the flow rate and the distance of the vial from the nozzle exit have been varied.

\section{Experimental Methods and techniques}

\begin{figure}
\includegraphics[width=0.45\textwidth]{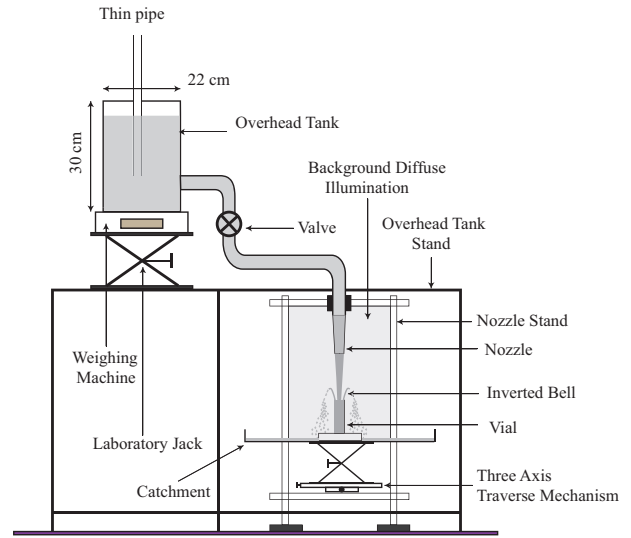}
\caption{\label{fig:setup} A schematic illustration of experimental apparatus}
\end{figure}

A schematic illustration of the experimental apparatus used for the present study is shown in figure \ref{fig:setup}. An overhead tank based on the Mariotte bottle principle is used to supply the water for the experiment at a constant flow rate. The Mariotte reservoir is placed on a precision weighing scale (Citizon CG 6102, readability or least-count 0.01 grams), which, in turn, is placed on a laboratory jack. A camera (IMPERX B2320M, 2352 x 1768 px) records the scale reading at 10 Hz during the experiment. The recorded scale images were further analyzed to get the mass flow rate. The overhead tank is connected to the nozzle via a flexible PVC pipe through a compressor ball valve. This valve and the laboratory jack are used to control the water flow rate.  

The supply reservoir and associated components are mounted on a rigid steel fixture. While the nozzle and receiving container are mounted on a separate heavy aluminum stand. The separation ensures that the nozzle and container alignment do not get disturbed due to the repeated filling of the supply reservoir. Fine cloth is used at the nozzle entrance to dampen any small fluctuations in the incoming water. The receiving container is mounted on a three-axis traverse and was used to align and control the separation between nozzle and receiver. The free-falling laminar jet stream at the low velocity was used as a reference to align the front camera. A twin-beam self-leveling cross-line laser (STANLEY CL90) was used to aid the alignment of the nozzle and container from the direction perpendicular to the line of sight of the front camera. At the time of performing experiments, the flow rate was adjusted such that the resulting bell shape has a distinct structure. Once the distinct flow structure was observed, the setup was left untouched to dampen any unsteady effects due to valve operation. After ensuring, by visual inspection, that the flow has reached a steady state, images of the water bell and weighing scale were acquired. This procedure was repeated until there were no further distinct flow features.  

The field of view for imaging has been illuminated with a diffused light produced by a 1000-W halogen lamp passing through a paper sheet in the background. Two cameras were used to record the formation and evolution of the water sheet structure at 1000 Hz. The front view was recorded by camera IDT OS10-4K: 3840 x 2400 px while the top view was recorded using the camera Nanosense MK-III, 1280 x 1024 px.

Experiments were performed for the cases presented in table \ref{tab:expPlan}. Where the parameter $Dn$, $Dv$, $H_{nv}$, $\dot{m}$, and $U_{ne}$  are nozzle internal diameter, vial internal diameter, the distance between nozzle exit and vial, mass flow rate and jet velocity at the nozzle exit. The theoretical jet velocity ($U_{th}$) and diameter ($D_{th}$) at vial mouth position have been calculated by applying Bernoulli's equation between nozzle exit and vial mouth. The length scale - \mygreen{$t_{g}$} and velocity scale - $U_g$ are calculated using the relations in equation \ref{eq:tg} and \ref{eq:ug} respectively.
\begin{equation}
t_g = \frac{D_v - D_{th}}{2}
\label{eq:tg}
\end{equation}

\begin{equation}
U_g =\frac{4\dot{m}}{\rho \pi (D_v ^2 - D_{th} .^2)} 
\label{eq:ug}
\end{equation}  

The non-dimensional parameter denoted by $Re$, $We$, and $Bo$ are Reynolds number ($\rho U^* D^* / \mu$ where $\rho$ and $\mu$ being the density and dynamic viscosity ), Webber number ($\rho U^{*2}D^*/\sigma$ where $\sigma$ being the surface tension), and Bond number ($D^*\sqrt{\rho g/2\sigma}$) respectively. Where, $U^*$ and $D^*$ are the characteristic velocity and diameter respectively. Two characteristic velocities :$U_{th}$, $U_{g}$ and diameters: $D_{th}$, \mygreen{$t_{g}$} are used to calculate $Re$, $We$, and $Bo$ and corresponding values are denoted with the suffix 'th' and 'g'. The properties of water used in the experiment are taken at $20 ^oC$.

\begin{table*}
\caption{\label{tab:expPlan} Geometric and flow parameters used in experiments}

\begin{ruledtabular}

\begin{center}
\begin{tabular}{ l c c c c c c c c c c} 

\multirow{2}{3em}{Case} & $D_n$  & $D_v$ & $H_{nv}$ & $\dot{m}$ & $D_{th}$ & $U_{ne}$ & $U_{th}$ & $Re_{th}$ & $We_{th}$ & $Bo_{th}$ \\
& (mm) & (mm) & (mm) &(g/s) & (mm) & (m/s) & (m/s) & & &\\ 
\hline

\multirow{3}{3em}{1} 
& 6 & 6 & 128.7& 14.98 & 3.37 & 0.53  & 1.67  & 5641 & 129.9  & 0.87 \\                                                  
& 6 & 6 & 128.7&16.74 & 3.54 & 0.59 & 1.69   & 6000  & 139.9  & 0.92 \\ 
& 6 & 6 & 128.7&23.68 & 4.10  & 0.84  & 1.79    & 7345  & 181.5  & 1.06 \\ 
\hline

\multirow{4}{3em}{2} 
 & 8 & 8 & 129.6& 28.02 & 4.55 & 0.56 & 1.72 & 7817  & 184.9 & 1.18 \\ 
 & 8 & 8 & 129.6&30.07 & 4.70 & 0.59 & 1.73 & 8131  & 193.9 & 1.22   \\ 
 & 8  & 8& 129.6& 34.58  & 4.99  & 0.68  & 1.76   & 8802 & 213.8 & 1.29 \\ 
 & 8 & 8 & 129.6&40.03 & 5.30  & 0.79 & 1.81   & 9590  & 238.9& 1.37  \\ 
 &8	&8	&129.6&53.07&	5.9&	1.06&	1.94&	11427&	304.9&	1.54\\

\hline

\multirow{5}{3em}{3} 
& 6  & 8&59.1  & 22.71  & 4.64  & 0.81  & 1.34 & 6222 & 115.0  & 1.20 \\ 
& 6 & 8&59.1 & 27.54  & 4.91 & 0.97  & 1.45   & 7124 & 142.4  & 1.27 \\ 
& 6  & 8&59.1  & 33.05  & 5.14  & 1.17  & 1.59    & 8165 & 178.7    & 1.33 \\ 
& 6  & 8&59.1   & 45.46 & 5.47  & 1.61  & 1.94    & 10568  & 281.6
& 1.41 \\ 
& 6 & 8&59.1 & 74.32  & 5.77  & 2.63 & 2.85   & 16375  & 640.8  
& 1.49  \\ 
\hline

\multirow{3}{3em}{4}
 & 6 & 6&31.8 & 10.46 & 3.91 & 0.37  & 0.87    & 3404      & 40.9   & 1.01 \\ 
& 6 & 6 &31.8 & 11.69 & 4.08 & 0.41 & 0.89   & 3637 & 44.6 & 1.06    \\ 
& 6 & 6 &31.8 & 20.81  & 4.95    & 0.74    & 1.08    & 5341  & 79.4       & 1.28  \\

\end{tabular} \\
\end{center}
\end{ruledtabular}
\end{table*}

\section{The differential equation of the water sheet}
The equations of motion for water bell were derived first by Boussinesq\cite{boussinesq1869theorie} and  were revisited by Lance and Perry\cite{lance1953water}. Later Taylor\cite{Taylor1959_a} non-dimensionalized the governing equations separating the constants that depended on gravity and pressure term. Clanet \cite{clanet2000stability} used these equations with a different characteristic length and velocity scale for non-dimensionalization. 

\begin{figure}
\includegraphics[width=0.45\textwidth]{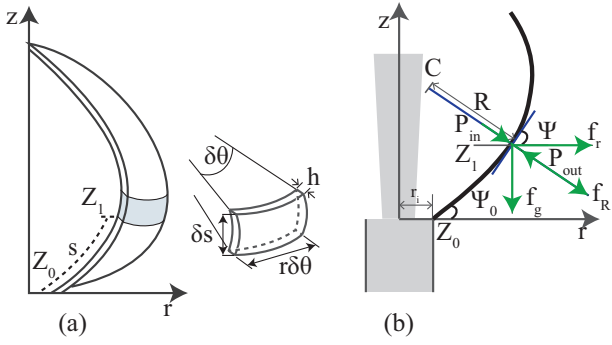}
\caption{\label{fig:fbd} A schematic showing (a) the coordinate system used and infinitesimal control volume (b) forces acting on the fluid element}
\end{figure}

However, for completeness the derivation is summarized again, following Lance and Perry \cite{lance1953water}, in the following passage. A three dimensional illustration of the open upward water bell is shown in figure \ref{fig:fbd} (a) along with the coordinate system. The water bell sheet is ejected from the vial edge at a point Z$_0$ with an initial ejection angle of $\psi_0$ and initial velocity of $v_0$. A differential fluid element is shown in the enlarged view at point Z$_1$, which is located at s arc length away from the Z$_0$. The differential element subtends an angle of $\delta \psi$ and $ \delta \theta$, with the z-axis, in the vertical and horizontal plane respectively. The thickness of the water sheet, h(s), can be calculated from conservation of mass and can be given as -   

\begin{equation}
h(s) =\frac{Q}{2 \pi r v(s) \rho} 
\label{eq:h}
\end{equation}

where $Q$ and $v(s)$ are the mass flow rate and fluid velocity respectively. 

The differential equation of the shape of water bell is derived considering an infinitesimal fluid element moving along a streamline with a velocity equal to the local fluid velocity at each point on the water sheet. The forces acting on the fluid element are shown in fig \ref{fig:fbd} (b) whereas $P_{in}$ and $P_{out}$ are the pressure on the  inside and outside of the water bell surface respectively; $f_r = \frac{2\sigma}{r} (\delta s \ r \delta \theta)$ and $f_R = = \frac{2\sigma}{R} (\delta s \ r \delta \theta)$ are the forces due to Laplace pressure as determined from the Young-Laplace equation\cite{rowlinson1982molecular} acting on the planes having a radius of curvature r and R whereas $\sigma$ denotes the surface tension of water-air interface; $f_g$ is the body force due to gravity. Applying the Reynolds transport theorem for non-inertial reference frame attached to the fluid element, having a differential mass ($\delta m = \rho \delta s \ r  \delta\theta \ h$), the conservation of linear momentum in the tangential and normal direction gives -
\begin{eqnarray}
- f_g \ sin \psi - \delta m \ v \frac{dv}{ds} = \nonumber \\
\frac{\partial}{\partial t} (\int_{\voldash} \rho \ v_{t} \ d \voldash) + \int_{A} v_{t} \ \rho (\ v_{t}\cdot n)  dA  \nonumber \\
- g \frac{dz}{ds} - v\frac{dv}{ds} = 0 \nonumber \\
v^2 = v_{0} ^2 - 2gz
\end{eqnarray}

where $\voldash$, $A$, $n$ are the volume, surface area and normal to the surface area of fluid element respectively.Tangential velocity with respect to the non-inertial frame of reference is denoted by $v_t$ ($=0$). 
\begin{eqnarray}
f_R + f_r sin\psi + (P_{in} - P_{out})(\delta s \ r \delta \theta) + \delta m \ g cos\psi - \delta m \frac{v^{2}}{R} \nonumber \\
= \frac{\partial}{\partial t} (\int_{\voldash} \rho \ v_{n} \ d \voldash) + \int_{A} v_{n} \ \rho (\ v_{n}\cdot n)  dA  \nonumber \\ \nonumber\\ \left(\frac{2 \sigma}{R} + \frac{2 \sigma}{r} sin\psi \right) + (P_{in} - P_{out}) + \rho h g cos\psi -\frac{v^2}{R} \rho h =0
\label{eq:normalMoment}
\end{eqnarray}

where $v_n (=0)$ is the normal component of velocity with respect to the non-inertial frame of reference. Using the definition of radius of curvature 
\begin{equation}
\frac{1}{R} = \frac{d\psi}{ds} = \frac{z''}{(1 + z'^{2})^{3/2}}
\end{equation}

where $z'$ and $z''$ are the first and second derivative with respect to r. The equation \ref{eq:normalMoment} can be transformed to

\begin{equation}
z'' = \frac{1 + z'^2}{e-\beta r}\left(\frac{\gamma}{e} +\beta z' + \alpha r (1 + z'^2)^{1/2} \right)
\label{eq:final}
\end{equation}

where $e= \sqrt{ 1 + \gamma z}$, $\gamma = g/v_0^{2}$,  $\alpha = 2 \pi (P_{in} - P_{out})/Q v_0$ and $\beta = 4\pi \sigma / Q v_0$. 

Equation \ref{eq:final} is solved using adaptive step size routine 'ode45' solver in MATLAB\textregistered \ using initial condition at vial edge ($z=0, r=r_i$, and $ dz/dr = tan (\psi_0)$). The 'ode45' routine is based on an explicit Runge-Kutta (4,5) formula,\cite{dormand1980family, shampine1997matlab}; It uses Runge-Kutta's $4^{th}$ and $5^{th}$-order accurate solution to calculate the error for the $4^{th}$ order solution.
\section{Results and Discussion}
\myred{A qualitative description of the observed phenomenon is presented first. Followed by the comparison of experimentally obtained bell shape profiles and theoretical prediction is presented. The scaling law identified in the present configuration is presented next. Finally, the variation of experimentally obtained bell diameter and height is presented.}

\subsection{A qualitative description}

\subsubsection{Case-I}
At the lower volume flow rate, \mygreen{initially} an inverted open water bell is formed for the case-I as shown in figure \ref{fig:case1}. At this stage the water sheet start to form at the center of the vial and rises and expands radially. Once the inverted water bell gains sufficient height, it takes a stable shape (figure \ref{fig:case1} (a)) {\color{black} (Multimedia view) }. The water sheet rolls back at the extremities and atomization breaks it into the droplets. The droplet generated due to atomization of the sheet, shed downward and do not imping on the inverted bell itself. On increasing the flow rate, \mygreen{for this particular case,} the inverted water bell diameter increases and the height starts to decrease (figure \ref{fig:case1} (b)).  The breakup occurs at the extremities and water droplets falls downward without disturbing the bell structure. However, the break up of the edge of the sheet is unstable and keeps on fluctuating. \mygreen{Finally,} \mygreen{The} sagging sheet comes into contact with the vial wall and traps the air inside making a widely studied water sheet structure - the classical water bell (figure \ref{fig:case1} (c)) {\color{black} (Multimedia view) } in this new arrangement. 

\subsubsection{Case-II}
The closed water bell was not observed to form in all the mass flow rates in case -II (figure \ref{fig:case2}). Spectacular, steady, inverted water bells can be observed in figures- \ref{fig:case2}(a) and \ref{fig:case2} (b) {\color{black} (Multimedia view) }. On further increasing the flow rate, the water sheet structure becomes horizontal, however the breakup of the sheet edge happens quite randomly making the sheet wiggle a little around the horizontal plane (figures \ref{fig:case2} (c)).  Further increase in the flow rate does not form a closed water bell, unlike the previous case. and the perimeter of the sheet grows. The breakup of the sheet is usually initiated by the random cusp formations. The umbrella type slope (figure \ref{fig:case2} (d)) of the water bell transforms to a constant slop when the flow rate is further increased (figure \ref{fig:case2} (e)).

\subsubsection{Case-III}
\begin{figure}
\includegraphics[width=0.45\textwidth]{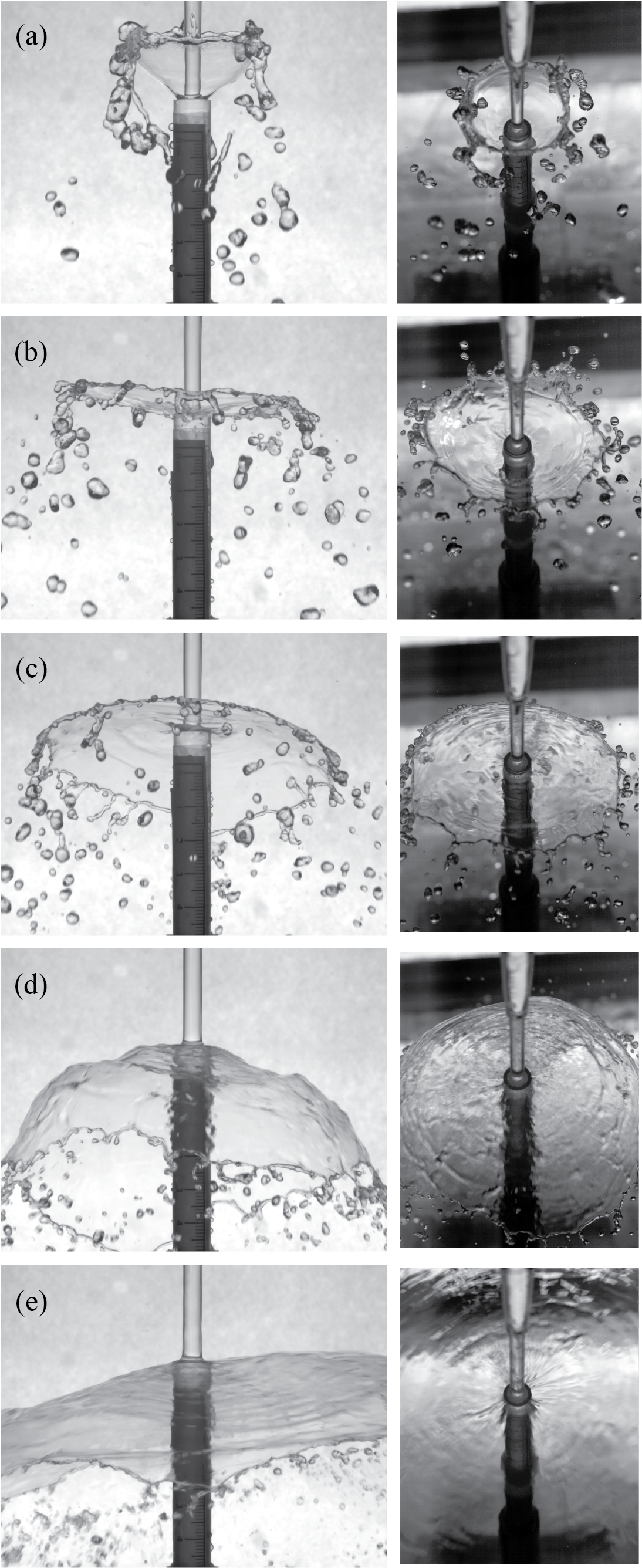}
\caption{\label{fig:case3} The transition of an inverted bell to open downward bell for case-III at flow rates (gram/sec) - (a) 22.71 (b) 27.54 {\color{black} (Multimedia view) } (c) 33.05 (d) 45.46 (e) 74.32; left column: front view and right column: isometric view}
\end{figure}

The formation of the flow structure for the case-III is presented in figure \ref{fig:case3}. An upward open bell could be seen in figure \ref{fig:case3} (a) at the lower flow rate with respect to this configuration. As the flow rate is increased, the shape of the bell transforms from slight upward (figure \ref{fig:case3}(b)) {\color{black} (Multimedia view) } to downward (figure \ref{fig:case3}(c)) water bell. The bell shape becomes brimmed downward as shown in figure figure \ref{fig:case3}(d). Finally, the slop of the downward open bell becomes straight when the flow rate is increased further (figure \ref{fig:case3}(d)).  

\subsubsection{Case-IV}
\begin{figure}
\includegraphics[width=0.42\textwidth]{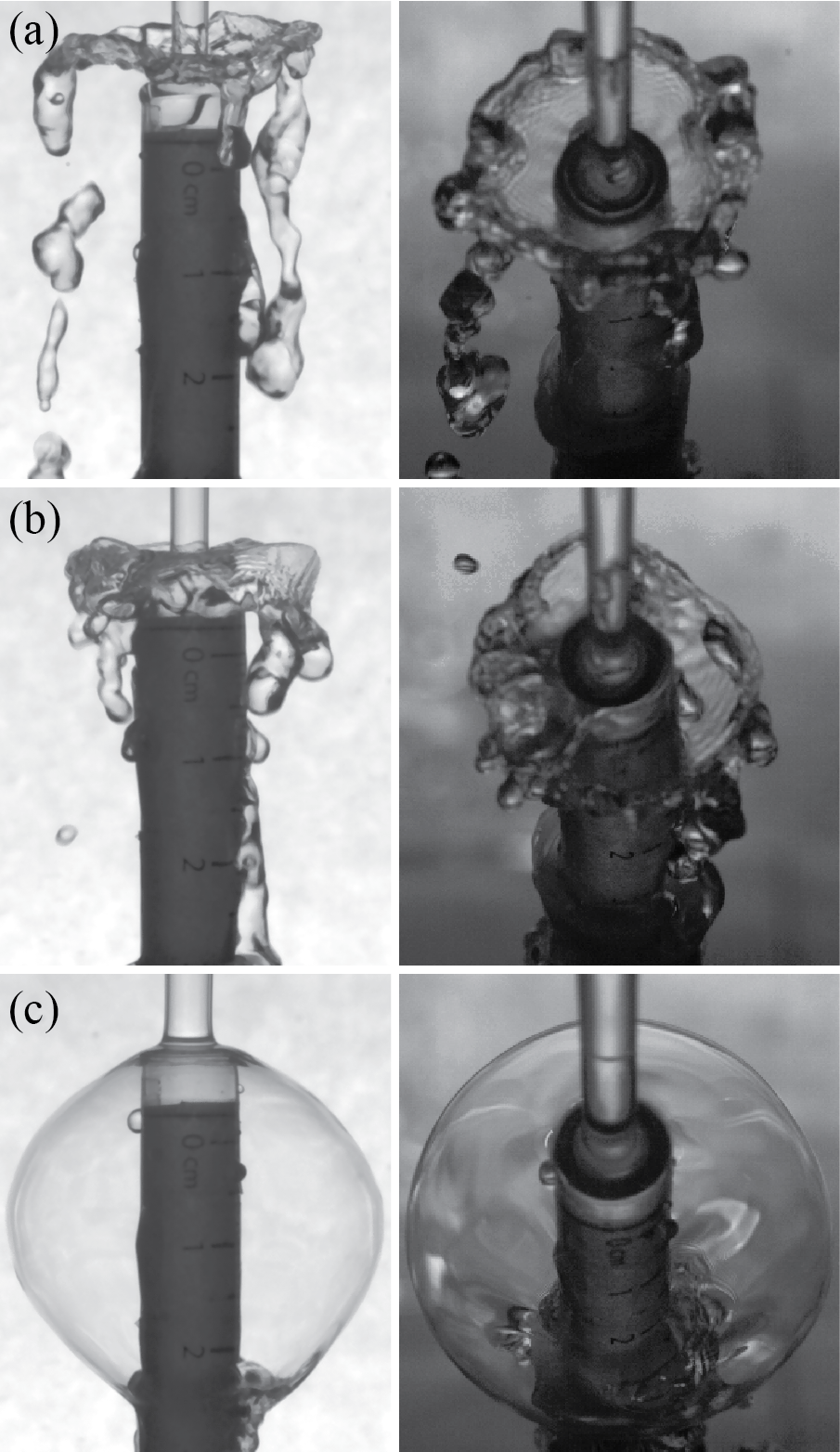}
\caption{\label{fig:case4} The transition of an inverted bell to bell for case-IV at flow rates (gram/sec) - (a) 10.46 (b) 11.69 (c) 20.81 {\color{black} (Multimedia view) }; left column: front view and right column: isometric view}
\end{figure}

In case-IV (figure \ref{fig:case4}) the formation of the closed bell is also observed. In this flow configuration, initially an upward inverted open bell forms (figure \ref{fig:case4}(a)), which flattens (figure \ref{fig:case4}(b)) on increasing the flow rate and comes into the vial wall contact due to its smaller size. On further increasing the flow rate it forms a small bell. However, when the jet flow into the vial is interrupted by a fast moving finger the new bell with a bigger diameter is formed. The interruption of the jet stream if repeated multiple times the diameter of the bell does not change further implying no hysteresis effects \cite{jameson2010water}. The bell formed at this stage is shown in figure \ref{fig:case4}(c) {\color{black} (Multimedia view) }. On further increasing the flow rate the size of the bell increases however no distinct flow structure were observed to be forming.

\graphicspath{ {figures/} }

\begin{figure}
\includegraphics[width=0.45\textwidth]{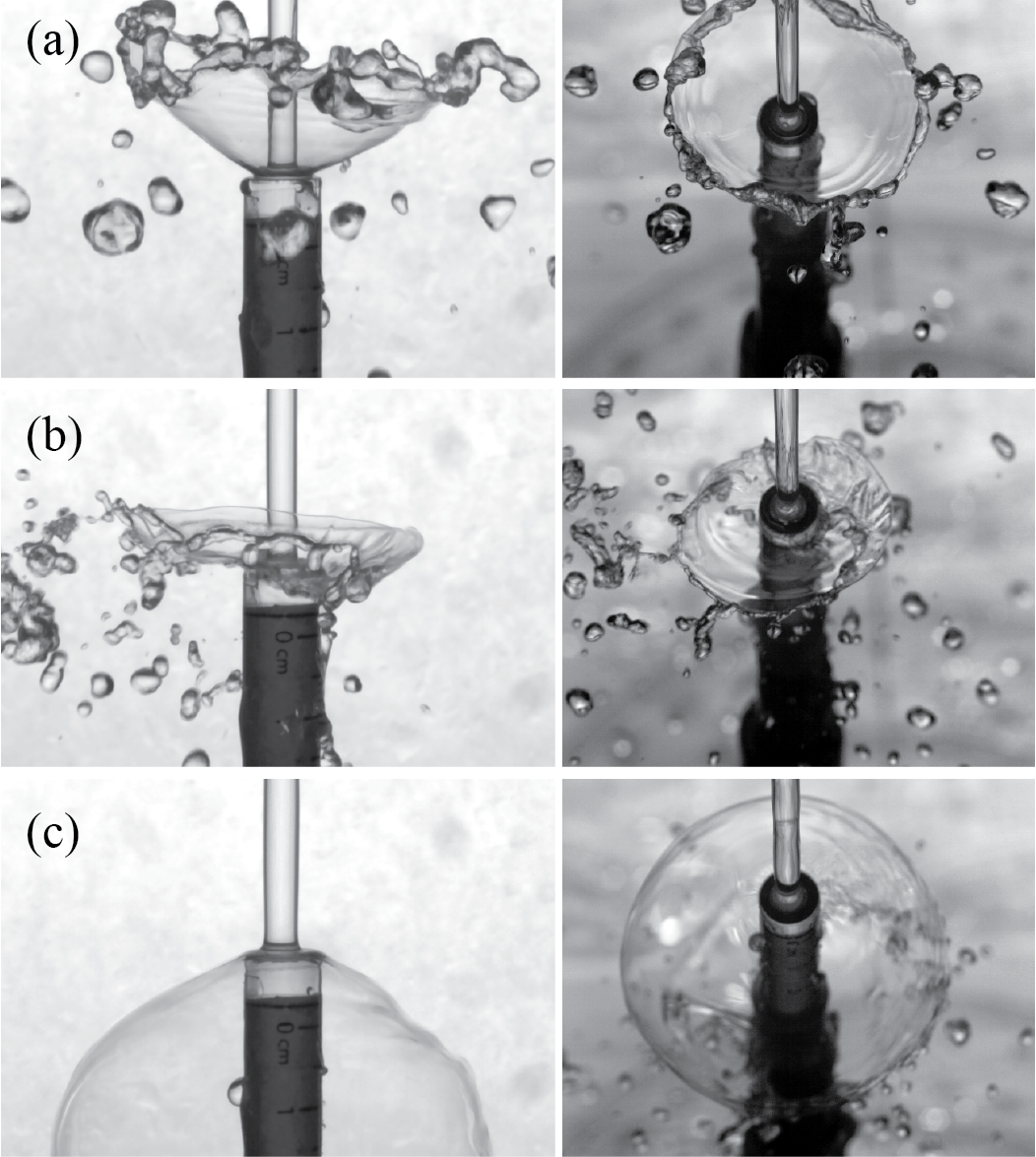}
\caption{\label{fig:case1} The transition of an inverted bell to bell for case-I at flow rates (gram/sec) - (a) 14.98 {\color{black} (Multimedia view) } (b) 16.74 (c) 23.68 {\color{black} (Multimedia view) }; left column: front view and right column: isometric view}
\end{figure}

\begin{figure}
\includegraphics[width=0.45\textwidth]{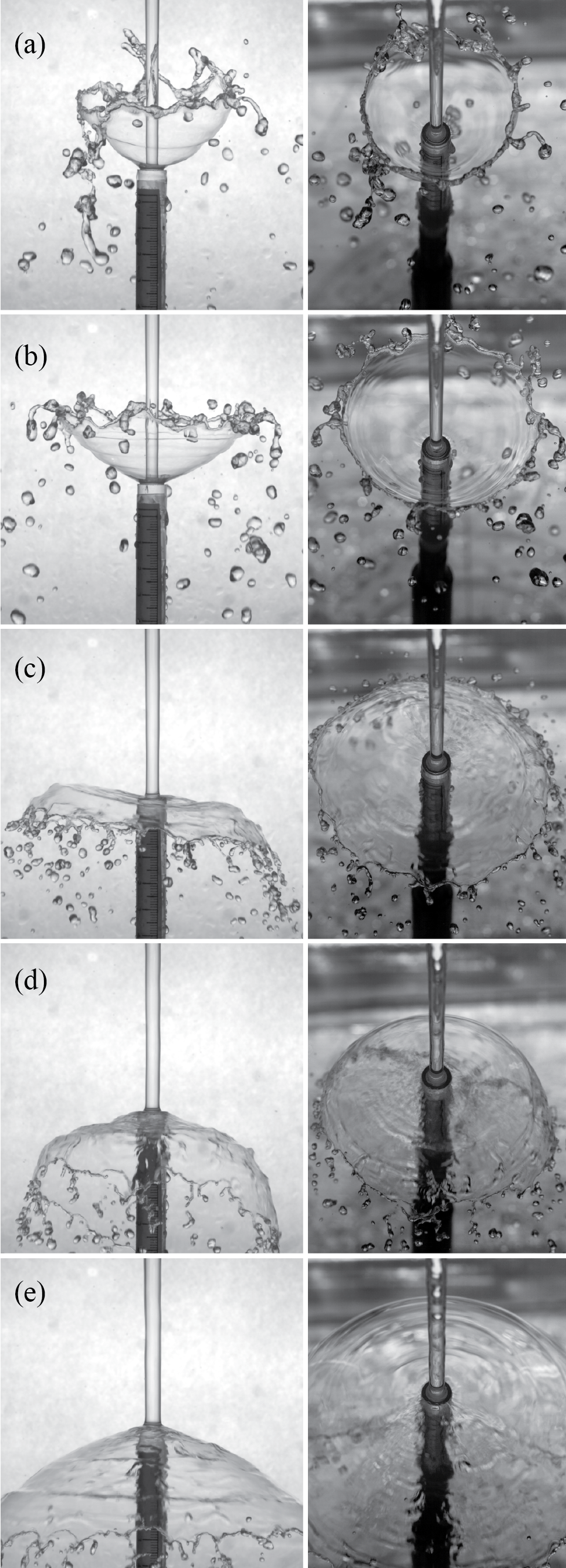}
\caption{\label{fig:case2} The transition of inverted bell to open downward bell for case-II at flow rates (gram/sec) - (a) 28.02 (b) 30.07 {\color{black} (Multimedia view) } (c) 34.58 (d) 40.03 (e) 53.07; left column: front view and right column: isometric view}
\end{figure}

\subsection{\myred{Comparison of bell shapes obtained from experiment and theoretical calculation}}
\begin{figure}
\includegraphics[width=0.45\textwidth]{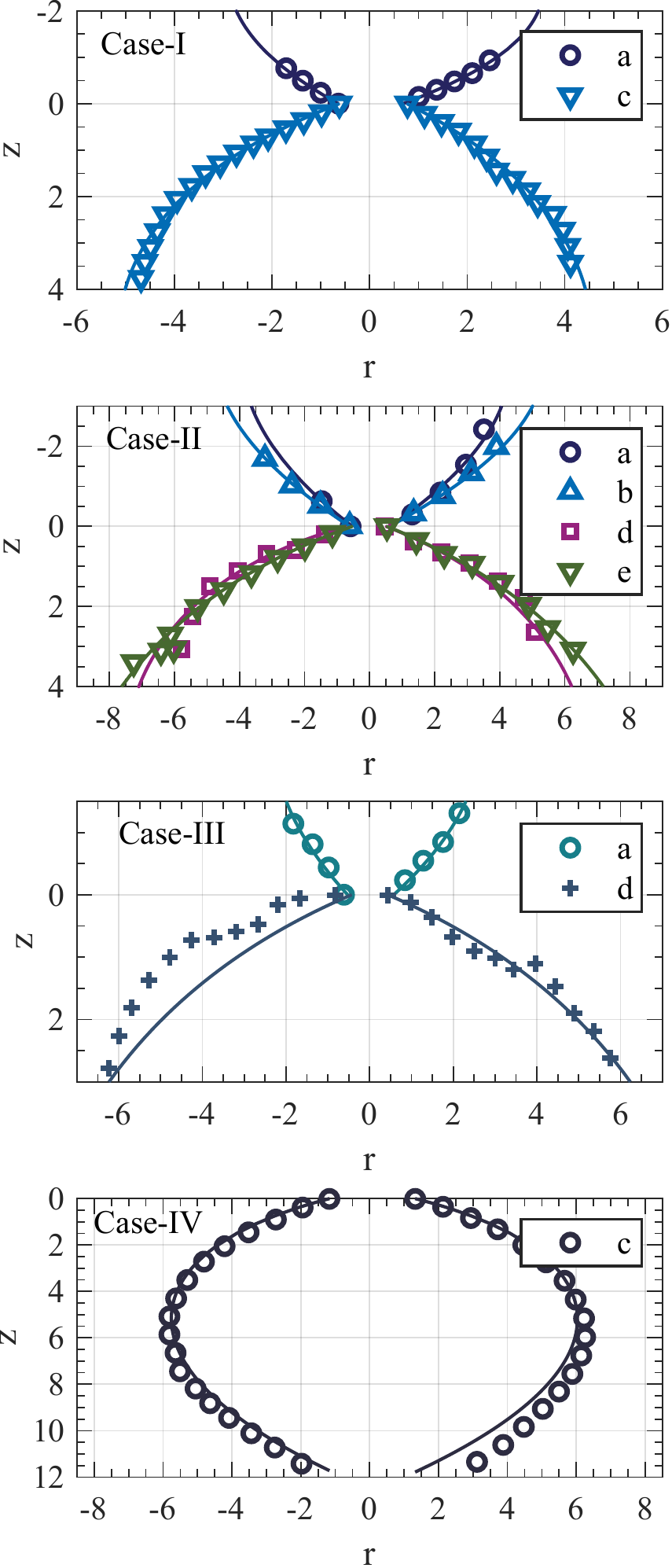}
\caption{\label{fig:bellProfl} Comparison of bell shapes obtained from theoretical calculation (denoted by solid lines) and extracted from experimental measurements (denoted by marker points); each subfigure shows the comparison for the case mentioned in top left corner and the legend shows the sub-cases}
\end{figure}

The governing equation as given in equation \ref{eq:final} is solved and the obtained profiles are compared with the shape extracted from the experimental images as shown in figure \ref{fig:bellProfl}. 

The mass flow rate, $\dot{m}$ and velocity, $U_{th}$ (see table \ref{tab:expPlan}) are used for $Q$ and $v_0$ in equation \ref{eq:final}. The values for $\sigma$, and $g$ are taken as 0.067 N/m and 9.8 m/s$^2$ respectively. The atmospheric pressure values are used for $P_{in}$ and $P_{out}$ for both open and close bells; though a small pressure difference exists in the case of closed bells. For the ejection angle, the angle as mentioned in table \ref{tab:gapTable} is used for the solution of the left and right branches of the bell profile. The left branch angle $\theta_L$ and right branch angle $\theta_{R}$ have a small asymmetry due to perturbations in the bell profile.  

The bell shape profile is extracted from the instantaneous snapshots as presented in the figures - \ref{fig:case1}, \ref{fig:case2}, \ref{fig:case3}, and \ref{fig:case4}. The images are binarized by selecting a suitable cut-off intensity value. This process separates the bell region from the image background. Finally, a boundary tracing is performed on the separated binary image to extract the edge of the bell. Note that separation of bell shape profile was not feasible in all the cases presented above due to view blockage created by the droplets. Hence the profile comparison is shown only for selected flow rates from each case. For case-I, the comparison is provided for the flow rates shown in figure \ref{fig:case1}, sub-figures (a) and c. Similarly, the case-II, case-III and case-IV show the profile comparison for the mass flow rate shown in figures \ref{fig:case2}, sub-figures (a), (b), (d), and (e); figure \ref{fig:case3}, sub-figures - (a) and (d), and figure \ref{fig:case4}, sub-figures - (c). A good agreement is obtained in case-I and case-II while the profiles in case-III and case-IV show some deviation due to error in ejection angle calculation which is supplied as an initial condition for the solution of equation  \ref{eq:final}. In comparison to the theoretical profiles, the experimentally obtained bells disintegrate prematurely due to the presence of instabilities.
    
\subsection{\myred{Scaling law governing the formation of open inverted, horizontal, and close bell formation} }
A comprehensive description of the non-dimensional parameters for each experimental case is presented in table \ref{tab:gapTable}. The parameters are calculated using the characteristic length scale (\mygreen{$t_g$}) and characteristic velocity scale ($U_g$) as defined in equation \ref{eq:tg} and equation \ref{eq:ug} respectively. In the rightmost column, a general classification of the observed bell shape is presented. Six categories of bell shapes have been assumed, approximately depending on the ejection angle and bell height - (a) open up: bell opening in the upward direction with $H_s > 11$ mm, (b) slightly up: bell opening up with height $5$ mm $< Hs < 11$ mm, (c) Horizontal: bell spreading horizontally with height $-5$ mm $< Hs < 5$ mm, (d) slightly down: bell opening down with height $-11$ mm $< Hs < -5$ mm, (e) open down - bell is opening down and its height $Hs < -11$ mm, and (f)close down - classical bell,closed downward, trapping air inside and its height is negative.

\begin{table*}
\caption{\label{tab:gapTable} Bell shape and non-dimensional parameters based on the gap characteristic scales}

\begin{ruledtabular}
\begin{center}
\begin{tabular}{ c c c c c c c c c c c} 

\multirow{2}{3em}{Case} & \mygreen{$t_g$} & $U_g$ & $Re_g$  & $We_g$ & $Bo_g$ & $\theta_L$ & $\theta_R$ & $H_s$ & $D_s$ & Bell Type \\

& (mm) & (m/s) &  &  &  & (deg) & (deg) & (mm) & (mm) & \\ 
\hline 

\multirow{3}{3em}{1} 
&1.31&	0.78&	1016&	10.9&	0.34&	30&  21.5& 13.7  &	39.0 & open up \\ 
&1.23&  0.91&   1115&    14	 &  0.32&  -&  -&  4.8	&   34.4 & horizontal \\ 
&0.95&	1.58&   1491&     32.3 &	0.25&   -22&  -28& -44.8&   53.4 & close down \\ 
\hline

\multirow{4}{3em}{2} 
&1.72&  0.83&	1418&	16.1&	0.45&	32&  29& 	25.4&	57.2& open up \\ 
&1.65&	0.92&   1505&	18.9&  0.43& 	28&  24& 	26.5&	71.8& open up \\
&1.5&   1.13&   1692 &    26.3&	0.39&	12.1&  -4.9&   -10.2&	81.3& slight down \\ 
&1.35&  1.43&   1912&    37.5&  0.35&	-13&  -17.5& 	-27.7&	90.6& open down \\ 
&1.05&  2.33&   2418 &    77.6&	0.27&	-16&  -18& -37.4	&	  -& open down\\
\hline

\multirow{5}{3em}{3}  

&1.68&	0.68&	1142&	10.7&	0.44&	36&  31.5& 	18&	    41.6 & open up\\
&1.54&	0.88&	1356&	16.4&	0.4	&   -&  15& 	    7.3&	 66.4& slight up\\
&1.43&	1.12&	1598&	24.7&	0.37&	-&  -&   -10.6&	81.7 & slight down\\
&1.26&	1.7	&   2146&	50.2&	0.33&	-10&  -16& 	-35.1&	104.9& open down\\
&1.11&	3.09&	3431&	145.8&	0.29&	-&  -14&    -35.3&	-    & open down\\

\hline

\multirow{3}{3em}{4}

&1.04&	0.65&	1173&	5.9&	0.27&	-&  43.9&  	4.6&	17.4& slight up\\
&0.96&	0.77&	1316&	7.8&	0.25&	-&  14&     0&	21.1& horizontal\\
&0.52&	2.32&	2612&	38.5&	0.14&	-21&  -20& 	-32.4&	36  & close down\\

\end{tabular} \\
\end{center}
\end{ruledtabular}
\end{table*}


\begin{figure}
\includegraphics[width=0.45\textwidth]{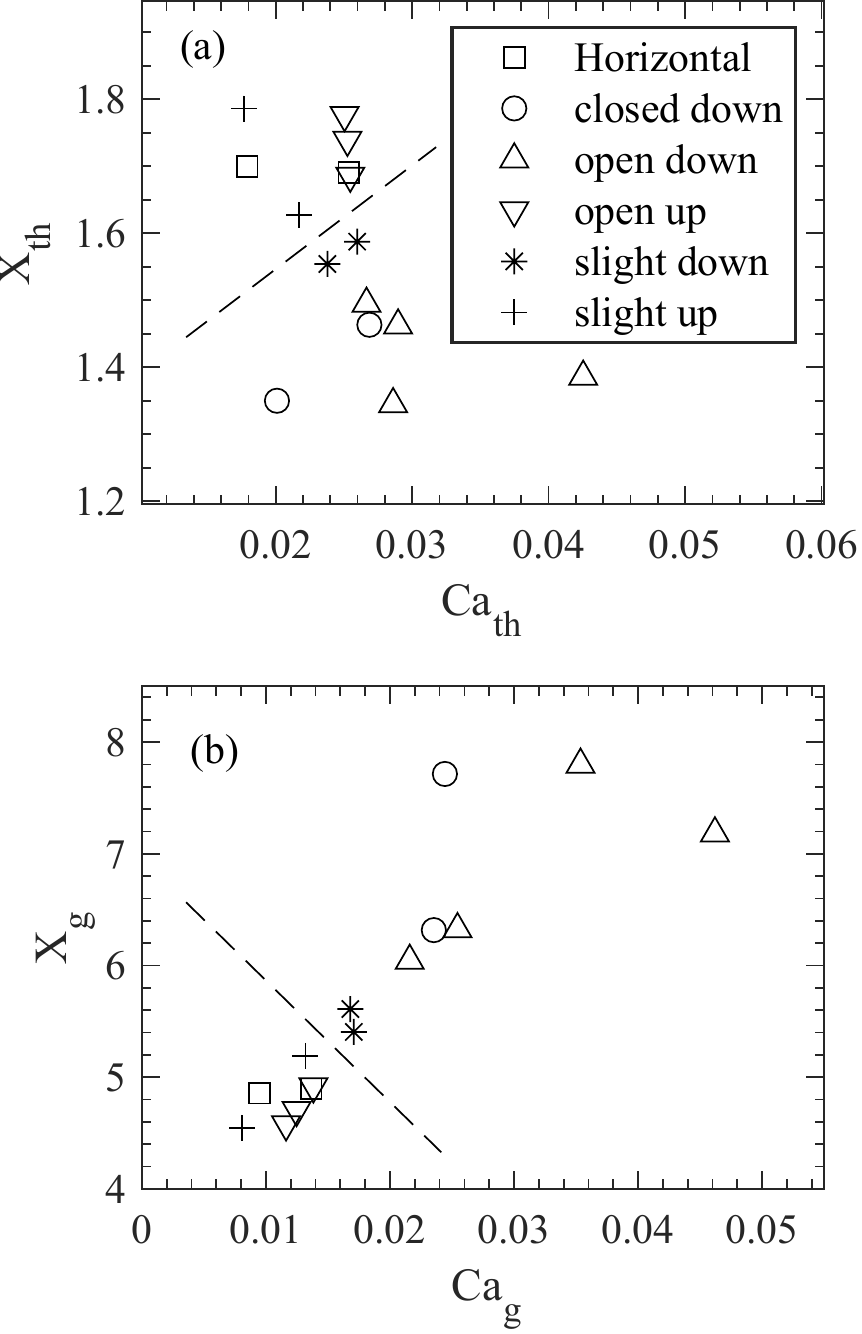}
\caption{\label{fig:scaling} Identified scaling law in the X and Ca plane, for the formation of particular kind of water bells, using velocity and length scale as (a) Uth and Dth and (b) Ugap and tgap}
\end{figure}

The identified scaling based on diameter ratio, $X_{th} = D_v / D_{th}$ and $X_{g} = D_v / t_{g}$, similar to the geometric diameter used by Clanet \cite{clanet2000stability} coupled with Capillary number\cite{rapp2016microfluidics} defined as $Ca_{th}=\mu U_{th} /\sigma $ and $Ca_{g}=\mu U_{g} /\sigma$ are presented in figure \ref{fig:scaling} (a) and \ref{fig:scaling} (b) respectively. The regimes corresponding to the formation of upward and downward bell profiles are separated by a dashed line in both the sub-figures. In figure \ref{fig:scaling} (a) the upward bell profiles are clustered in the top left corner whereas the downward bell profiles are getting placed below the dashed line. The upward bell shapes are clustered in the bottom left corner and downward bell shapes are situated on the other side of the dashed line in figure \ref{fig:scaling} (b). Hence, the presented scaling predicts the formation of the upward and downward bells with high accuracy in the $X_{th}$ and $Ca_{th}$ and $X_{g}$ and $Ca_{g}$ plane. However, the scaling does not provide a provision for any distinction in the formation of closed downward and open downward bells with the existing data set. Further experiments are being performed to get further insights into this regime.

\subsection{Variation of the bell diameter and height}
\begin{figure}
\includegraphics[width=0.45\textwidth]{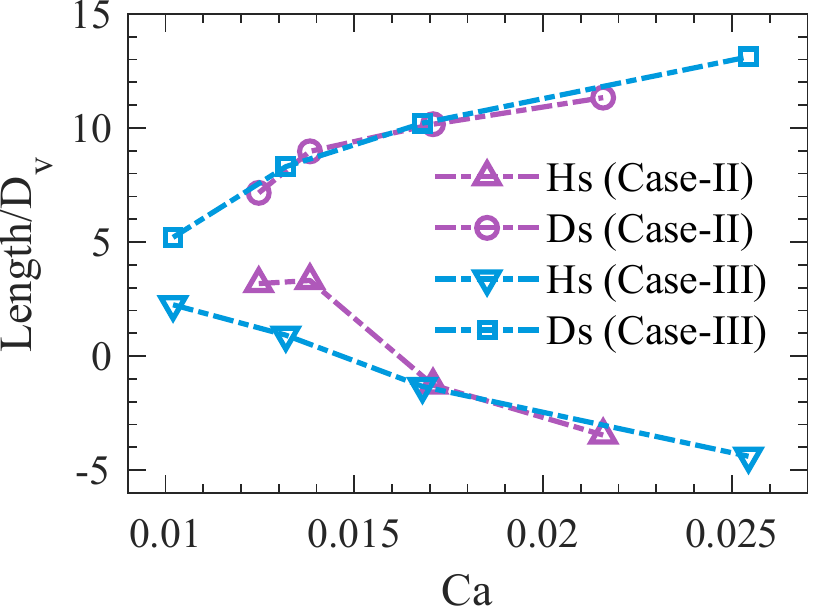}
\caption{\label{fig:case2n3HsDs} \mygreen{Variation} of non-dimensionalized bell diameter ($D_s$) and height ($H_s$) with Ca$_g$ based on the characteristic velocity scale $U_{g}$ for case-2 (figure \ref{fig:case2}) and case - 3 (figure \ref{fig:case3}) }
\end{figure}

The bell shape profiles, near the root, can be correctly determined using the governing equation (equation \ref{eq:final}). However, the shapes disintegrate prematurely as confirmed by the experimental observations. To find a scaling relation in the pattern of disintegration the bell diameter and its height are plotted with different non-dimensionalized numbers. Here we present the variation of bell diameter ($Ds$) and its height ($Hs$) with respect to the mass flow rate. The height is taken as the distance between the vial mouth to the outermost edge of the bell with positive in the upward direction. The measurements are taken manually by selecting the reference distances in the images.  The general trend of the diameter and height variation for water bell structure formed in case-II (figure \ref{fig:case2}) and case-III (figure \ref{fig:case3}) are presented in figure \ref{fig:case2n3HsDs}. The bell diameter and heights presented in this figure are non-dimensionalized with respect to the vial diameter $D_v$. The diameter and height show a strong dependence on the capillary number. A good collapse of the data is obtained. This curve can be used to predict the height and diameter of the water bells. Note, that the measurement of the bell diameter, height, and angle of ejected sheet near the vial mouth was not possible for all the cases. Therefore, the trend analysis plots are only taken for the cases which have sufficient number of data points. However, the full list of measured data points are presented in table \ref{tab:gapTable}.     

\section{Conclusion}
In the present work, a new configuration is considered for water bell formation where we have investigated the formation of water sheet structures during the impingement of vertically downward jet on a vial filled with liquid. This investigation is inspired by the serendipitous observation made during the washing of a glass vial in the kitchen sink.

The water sheet structures for four different geometric configurations are considered where, nozzle diameter, vial diameter and the separation between nozzle and vial mouth are varied. For each configuration, mass flow rate is changed from low to high and the distinct shape of sheet structures observed.

    For a given geometric configuration, at low flow rate the water sheet structure forms a spectacular steady inverted bell that transforms to a horizontal water sheet followed by either a downward open bell or even a classical closed water bell. The formation of the classical water bells has been only reported by use of a deflector disk or by head on collision of two jets in earlier studies. The steady inverted bell to downward classical bell formation in the present study opens up the possibility of a general frame work of formation of water sheet structures for different $We, Bo $ and $Re$ in future studies. 

\myred{Identified scaling law predicts the formation of upward and downward bell shapes in the diameter ratio X and capillary number Ca plane. The variation of experimental bell shape height and diameter also shows a strong dependence on the capillary number. }

\begin{acknowledgments}
We wish to acknowledge the support of Department of Science and Technology (DST) India for providing the high speed cameras in other grants. 

\end{acknowledgments}

\section*{Data Availability Statement}

The data that support the findings of this study are available within the article.

\clearpage
\nocite{*}
\bibliography{aipsamp_final}
\end{document}